\documentclass[useAMS,usenatbib]{mn2e}
\usepackage{graphicx}
\usepackage{amsmath,amsfonts,amssymb}
\usepackage{wrapfig}
\usepackage{epsfig}
\usepackage{psfrag}
\usepackage{subfigure}
\usepackage{pstricks,pst-plot} 

\begin{document}

\title[Bursts in Parkes Multibeam Data]
{A search for dispersed radio bursts in archival Parkes Multibeam Pulsar Survey data}

\author[Bagchi, Cortes \& McLaughlin]
{\parbox[t]{\textwidth}{Manjari Bagchi$^{1}$\thanks{Email: Manjari.Bagchi@mail.wvu.edu}, Angela Cortes Nieves$^{1}$ and Maura McLaughlin$^{1,2}$}\\
\vspace*{3pt} \\
\\ $^1$ Department of Physics, White Hall, West Virginia University, Morgantown, WV 26506, USA
\\ $^2$ Also adjunct at NRAO, Green Bank Observatory, PO Box 2, Green Bank, WV 24944, USA
}
\maketitle

\begin{abstract}

A number of different classes of potentially extra-terrestrial bursts of radio emission have been observed in surveys with the Parkes 64m radio telescope, including ``Rotating Radio Transients'', the ``Lorimer burst'' and ``perytons''. Rotating Radio Transients are radio pulsars which are best detectable in single-pulse searches. The Lorimer burst is a highly dispersed isolated radio burst with properties suggestive of extragalactic origin. Perytons share the frequency-swept nature of the Rotating Radio Transients and Lorimer burst, but unlike these events appear in all thirteen beams of the Parkes Multibeam receiver and are probably a form of peculiar radio frequency interference. In order to constrain these and other radio source populations further, we searched the archival Parkes Multibeam Pulsar Survey data for events similar to any of these. We did not find any new Rotating Radio Transients or bursts like the Lorimer burst. We did, however, discover four peryton-like events. Similar to the perytons, these four bursts are highly dispersed, detected in all thirteen beams of the Parkes multibeam receiver, and have pulse widths between 20--30 ms. Unlike perytons, these bursts are not associated with atmospheric events like rain or lightning. These facts may indicate that lightning was not responsible for the peryton phenomenon. Moreover, the lack of highly dispersed celestial signals is the evidence that the Lorimer burst is unlikely to belong to a cosmological source population.

\end{abstract}



\section{Introduction} 
\label{lb:introductions} 

After the discovery of ``Rotating Radio Transients'' (RRATs) by \citet{mll06} through single-pulse searches,  more searches for non-repetitive, dispersed signals in radio pulsar survey data have led to discoveries of more RRATs and other signals. The new discoveries have established that RRATs are not a distinct neutron star population but are extreme examples of the radio pulsar population, and their pulsed emissions are either highly modulated or sufficiently infrequent which makes them easier to find in single-pulse studies \citep{wsrw06, bb10, kklsm11}. One intriguing discovery was a very bright ($30$ Jy) dispersed (dispersion measure of $375~{\rm cm^{-3} \, pc}$ ) radio burst by \citet{lbmnc07} while re-analyzing archival data of a survey of the Magellanic Clouds using the multibeam receiver on the 64-m Parkes radio telescope in Australia. This burst, commonly known as the ``Lorimer burst'', was detected in three adjacent beams. As the burst followed the dispersion delay law ($\delta \, t \propto \nu^{-2}$) of cold plasma and showed pulse width broadening ($w \propto \nu^{-4}$) due to interstellar scattering, it was believed to be of celestial origin. The high value of the dispersion measure (DM) indicated that the burst was extragalactic, with a distance estimate of 500$-$1000 Mpc using the latest models for Galactic and extragalactic electron density \citep{cl02, io03, in04}. Unfortunately, the true origin of this burst is not yet understood, although there are various suggestions \citep{pp07, vach08, pp10, ep09}.

The above discovery prompted \citet{bbemc11} to search for similar bursts in four other archival pulsar survey data sets, taken over the years 1998$-$2003 with the same multibeam receiver on the Parkes telescope. As a result, they detected 16 bursts, 14 of which had DM values in the range of $350-406~{\rm cm^{-3} \, pc}$  and two with DM values of 220 and 216 $~{\rm cm^{-3} \, pc}$. They detected 15 out of these 16 bursts in all 13 beams of the receiver. Twelve bursts occurred on 23 June 1998. Of the other four, one occurred on 25 June 1998, one on 1 March 2002, one on 30 June 2002, and one on 2 July 2003. It is quite interesting that 15 out of 16 bursts occurred either in late June or in early July (Australian mid-winter). Fourteen of these bursts occurred on rainy days \textit{e.g.} on 23 June 1998, total rain was 17.2 mm and average wind speed was 24 km/h, on 25 June 1998, total rain was 5.4 mm and average wind speed was 11 km/h, on 1 March 2002, there was no rain and the average wind speed was 14 km/h, there was no rain on 30 June and the average wind speed was 9 km/h, on 2 July 2003, there was total rain of 0.8 mm and the wind speed was 5 km/h \footnote{The weather information at the Parkes telescope site has been checked from \textit{http://www.parkes.atnf.csiro.au/cgi-bin/monitoring/wstats.cgi}.}.

All these bursts occurred around midday. Although these bursts in general follow the dispersion law of cold plasma, there are some significant deviations. Moreover, these bursts are not equally strong over the entire frequency band and are much brighter and broader ($30 - 50$ ms) than the Lorimer burst. These facts suggested a terrestrial origin. These peculiar characteristics inspired the discoverers to name these bursts  ``perytons'' after the fictional character with the same name which looks like a winged deer but has the shadow of a man. This discovery has created some suspicion as to whether the Lorimer burst was also of terrestrial origin. 

Recently, five more perytons which occurred a few minutes prior to the first peryton of \citet{bbemc11}, have been reported by \citet{kbb12}. There have been discoveries of other possible extragalactic radio signals too. One is the 7 mJy signal at a DM of 745 ${\rm cm^{-3} \, pc}$ found in archival Parkes Multibeam Pulsar Survey (PMPS) data \citep{kklsm11}. The other example is a radio burst observed at a frequency of 328 MHz in M31 using the Westerbork Radio Telescope  in the Netherlands \citep{eduardo}. The DMs of these bursts are consistent with either Galactic or extragalactic signals, making their origins unclear.

Clearly, discovery of more highly dispersed radio bursts will enable us to better understand their nature. That is why we decided to search the archival PMPS data for such objects. In our analysis we found four new examples of highly-dispersed single pulses occurring in all 13 beams of the multibeam receiver, with similar characteristics to the perytons described above.

The rest of the paper is organized as follows. In Section \ref{sec:data} we briefly describe the PMPS and the resulting data. We report our search algorithm and results in Section \ref{sec:results}. We extend our discussions about the bursts in Section \ref{sec:disc}. Finally we present our conclusions in Section \ref{sec:conclu}.

\section{Data and Analysis}
\label{sec:data} 

The PMPS was a  search for pulsars in the region $\mid b \mid < 5^{\circ}$ and $260^{\circ} < l< 50^{\circ} $ using the multibeam receiver of the 64-m Parkes radio telescope. This survey began in August 1997 and ended in October 2003. Receivers for each of 13 beams are sensitive to two orthogonal linear polarizations. Signals from each polarization of each beam are detected in 96 filters, each 3 MHz wide (total bandwidth 288 MHz, central frequency 1374 MHz), upon which they are added in polarization pairs, high-pass filtered with a cut-off of 0.2 Hz, and integrated for 250 $\mu {\rm s}$. Data had been collected by interleaving pointings on a hexagonal grid, resulting in complete sky coverage with adjacent beams overlapping at half-power points. Each pointing covers an area about 0.6 ${\deg}^{2}$, resulting in sky coverage at a rate of 1 ${\deg}^{2}/{\rm hr}$. The total survey area comprised about 3100 pointings. This extremely successful survey has yielded around 800 pulsars so far \citep{pks4}. See \citet{pks1}, \citet{pks2}, and \citet{pks3} for more details about the survey and the multibeam receiver.

In the present work, we have used the PMPS data where all time series were first de-dispersed for a number of trial values of DMs. The time series were smoothed by convolution with boxcars of various widths to increase sensitivity to broadened pulses, with a maximum boxcar width of 32 ms. Because the optimal sensitivity is achieved when the smoothing window width equals the burst width, the sensitivity is lower for burst durations greater than 32 ms. For each of these time series, mean and root-mean-square deviations of the noisy time series were calculated and signals with peak signal-to-noise about some level were recorded 
\citep[see details in][]{cm03}. 

\section{Results} 
\label{sec:results}

We searched for non-repetitive signals having SNR greater than 7 in the DM range 200$-$2000 ${\rm cm^{-3} \, pc}$. Many of the known RRATs and the high DM burst reported by \citet{kklsm11} are detected. These objects in general have SNR $> 7$ only in one beam. We did not find any new Lorimer burst-like events (\textit{i.e.} strong signals in one or a few beams showing $\delta \, t \propto \nu^{-2}$ relation in the frequency - time plots). The data contains a large amount of radio frequency interference (RFI). 
Most of the time, the RFI has peak SNR at zero DM and in other times, it shows high SNR over a very wide range of DM in single-pulse plots, but does not show the $\delta \, t \propto \nu^{-2}$ relation in the frequency - time plots. Moreover, even when a particular RFI signal shows up in multiple beams, the DM corresponding to the peak SNR is significantly different for different beams.  

We found four quite distinctive bursts. All of these bursts have high SNR in all thirteen beams. One burst occurred on 3 August 1998, two on 12 August 1998, and another on 15 September 2003. The two bursts on 12th August 1998 are separated by around $1.5$ min. The details of these bursts are given in Table \ref{tb:peryton1}. These bursts were detected in all 13 beams with pulse width between $\sim 20 - 30$ ms similar to the perytons. Moreover, the DM values for these bursts are similar to that of the majority of the perytons, i.e. between $350 - 430$ ${\rm cm^{-3} \, pc}$. Another similarity is that these bursts also happened during daytime (between 12 -- 14 AEST). 

\begin{figure}
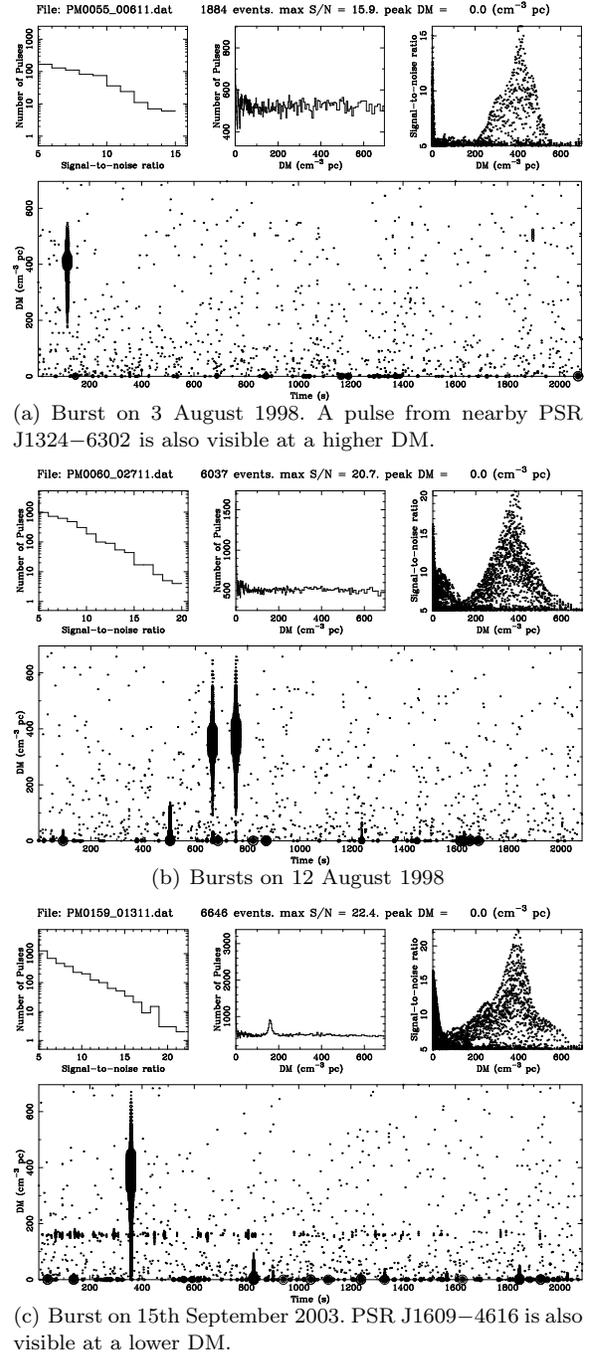

 \begin{center}
\subfigure[Burst on 3 August 1998. A pulse from nearby PSR J1324$-$6302 is also visible at a higher DM.]{\label{fig:singlepulse_sub1}\includegraphics[width=0.3\textwidth,angle=270]{f1a.ps}}
\subfigure[Bursts on 12 August 1998]{\label{fig:singlepulse_sub2}\includegraphics[width=0.3\textwidth,angle=270]{f1b.ps}}
\subfigure[Burst on 15th September 2003. PSR J1609$-$4616 is also visible at a lower DM.]{\label{fig:singlepulse_sub3}\includegraphics[width=0.3\textwidth,angle=270]{f1c.ps}}
  \end{center}
\caption{Single-pulse plots for the central beams for all of the bursts. Each burst detected above a threshold of 5 sigma is shown. The plots are (top left) number of pulses vs signal-to-noise, (top middle) number of pulses vs. DM, and (top right) signal-to-noise vs DM. The bottom plot shows the DM vs time for all bursts, where the size of the
circle is proportional to the signal-to-noise of the burst. The signals at zero DM in the top-right panels and the bottom panels are due to ground-based RFI. They are clearly distinguishable from signals of astrophysical nature due to their non-dispersed nature.}
\label{fig:all_singlepulse}
\end{figure}

\begin{figure*}
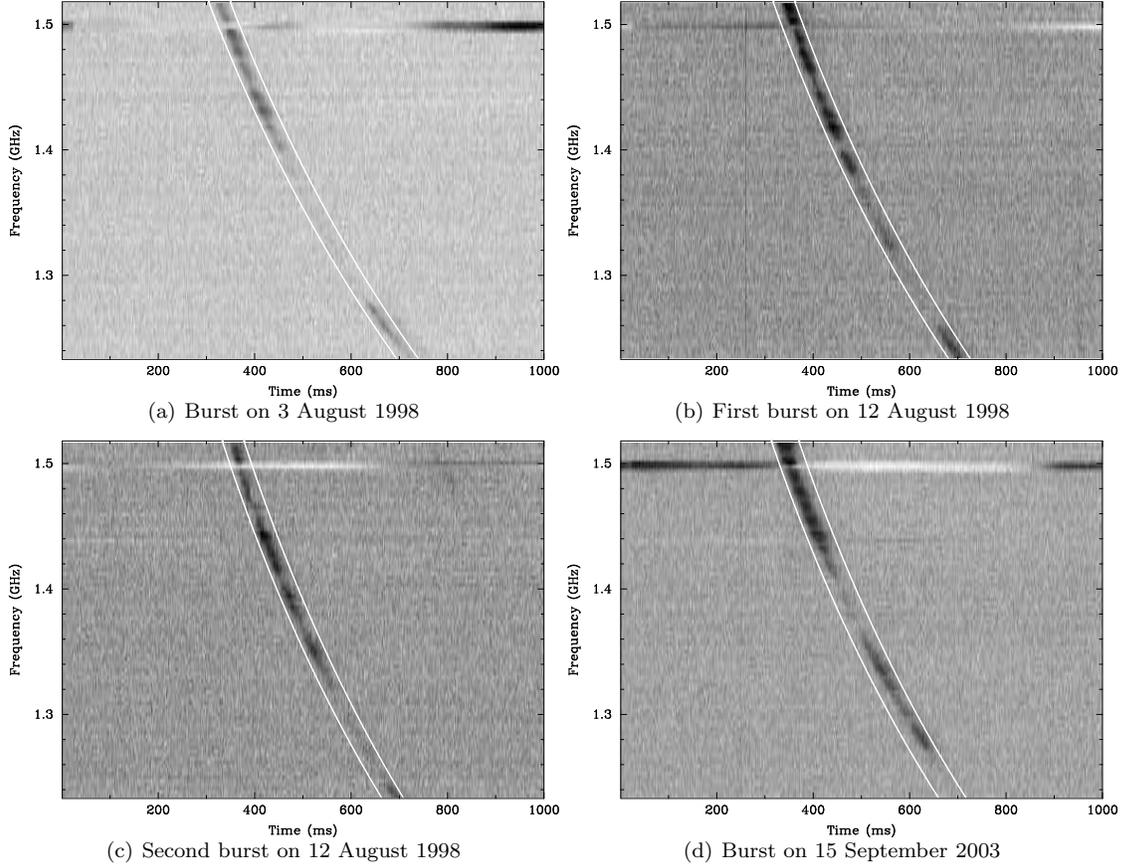

 \begin{center}
\subfigure[Burst on 3 August 1998]{\label{fig:ddfreqsub1}\includegraphics[width=0.3\textwidth,angle=270]{f2a.ps}}
\subfigure[First burst on 12 August 1998]{\label{fig:ddfreqsub2}\includegraphics[width=0.3\textwidth,angle=270]{f2b.ps}}
\subfigure[Second burst on 12 August 1998]{\label{fig:ddfreqsub3}\includegraphics[width=0.3\textwidth,angle=270]{f2c.ps}}
\subfigure[Burst on 15 September 2003]{\label{fig:subfig3}\includegraphics[width=0.3\textwidth,angle=270]{f2d.ps}}
 \end{center}
\caption{Dispersion delay across the entire frequency band for the bursts mentioned in Table \ref{tb:peryton1}. For each case, the signals in all beams have been summed. The white lines represent the dispersion delay law ($\delta \, t \propto \nu^{-2}$) for cold plasma. In each case, RFI at 1.5 GHz is present.}
\label{fig:allfreqtime}
\end{figure*}

\begin{figure*}
 \begin{center}
\subfigure[Burst on 3 August 1998]{\label{fig:pulseshape_sub1}\includegraphics[width=0.4\textwidth,angle=0]{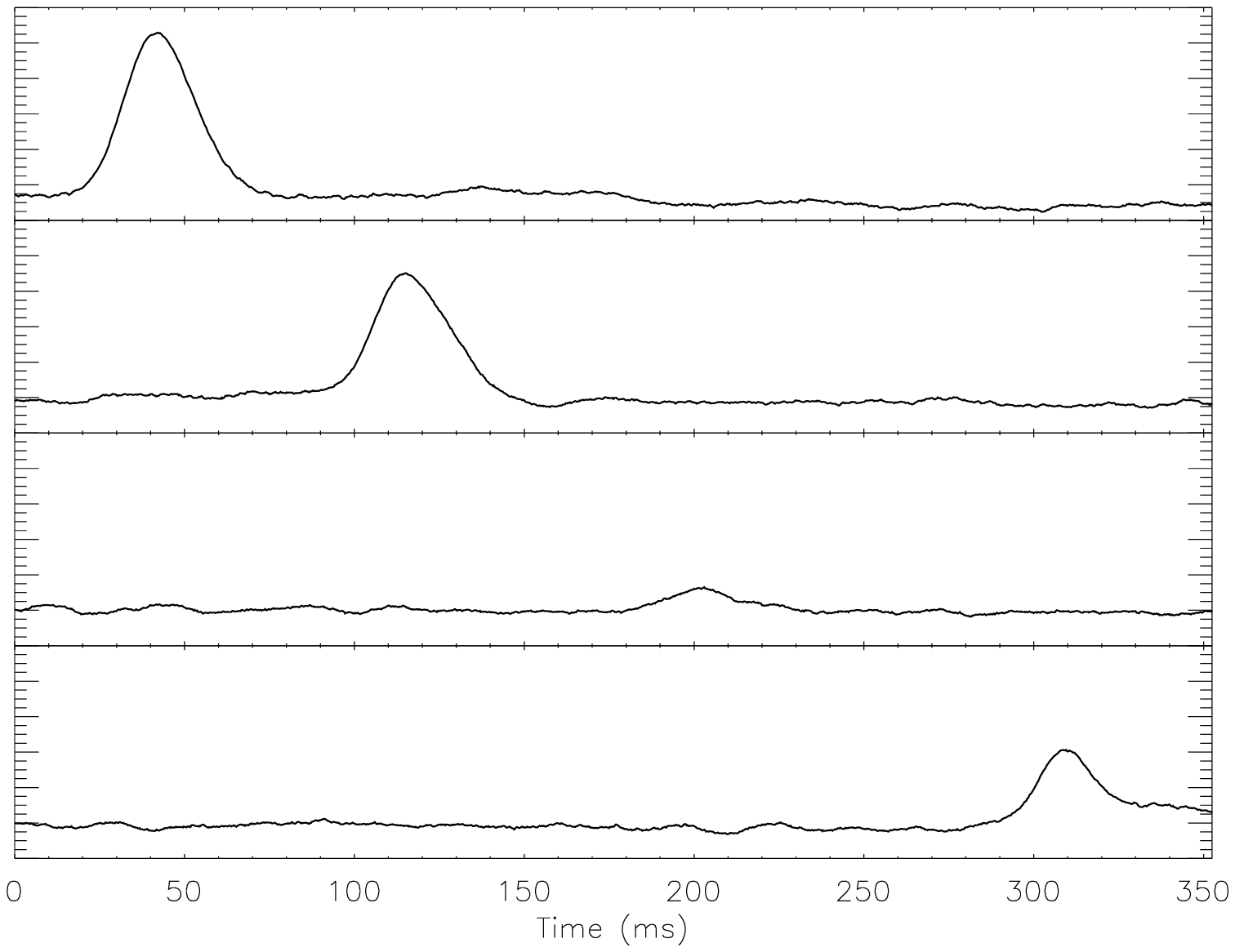}}
\subfigure[First burst on 12 August 1998]{\label{fig:pulseshape_sub2}\includegraphics[width=0.4\textwidth,angle=0]{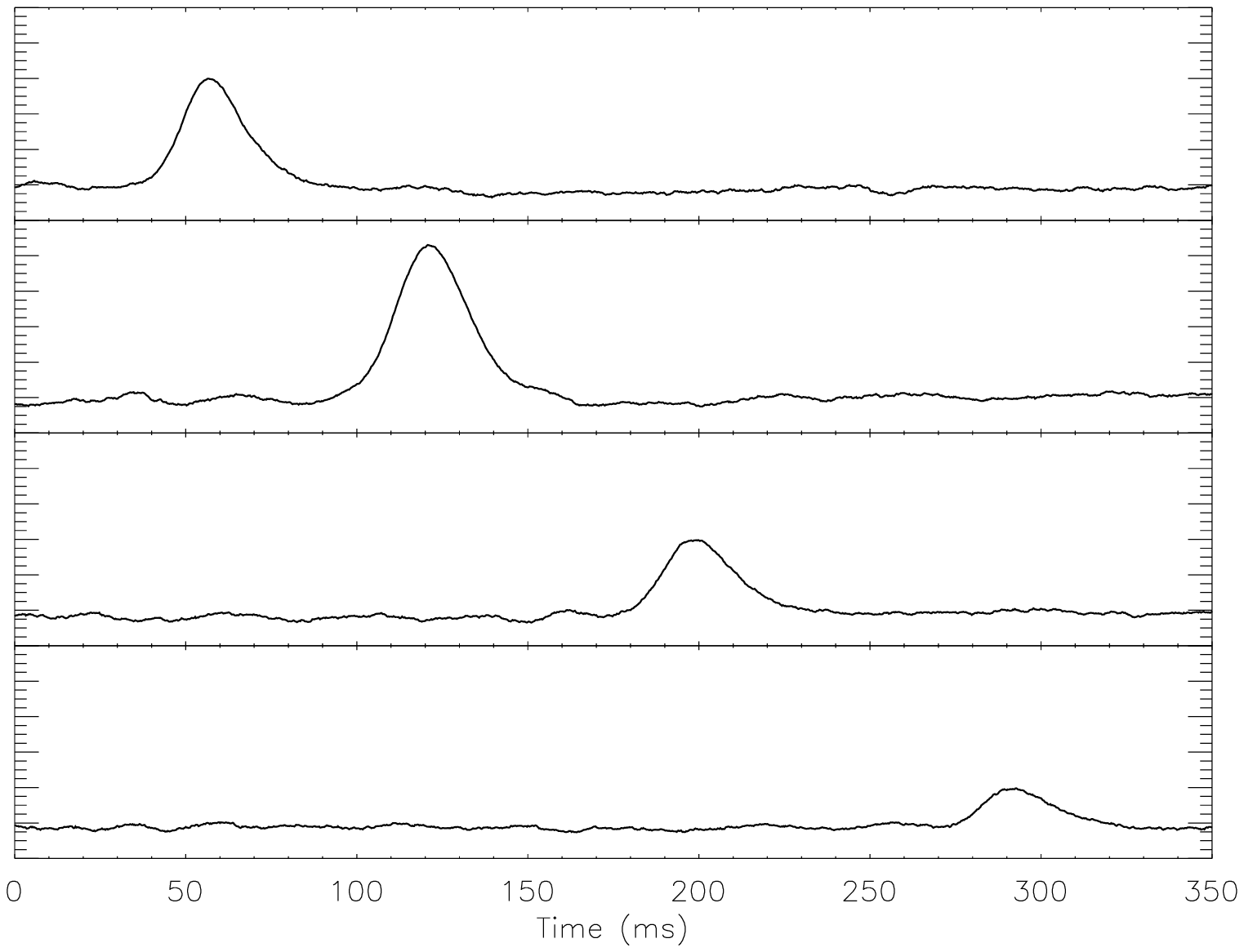}}
\subfigure[Second burst on 12 August 1998]{\label{fig:pulseshape_sub3}\includegraphics[width=0.4\textwidth,angle=0]{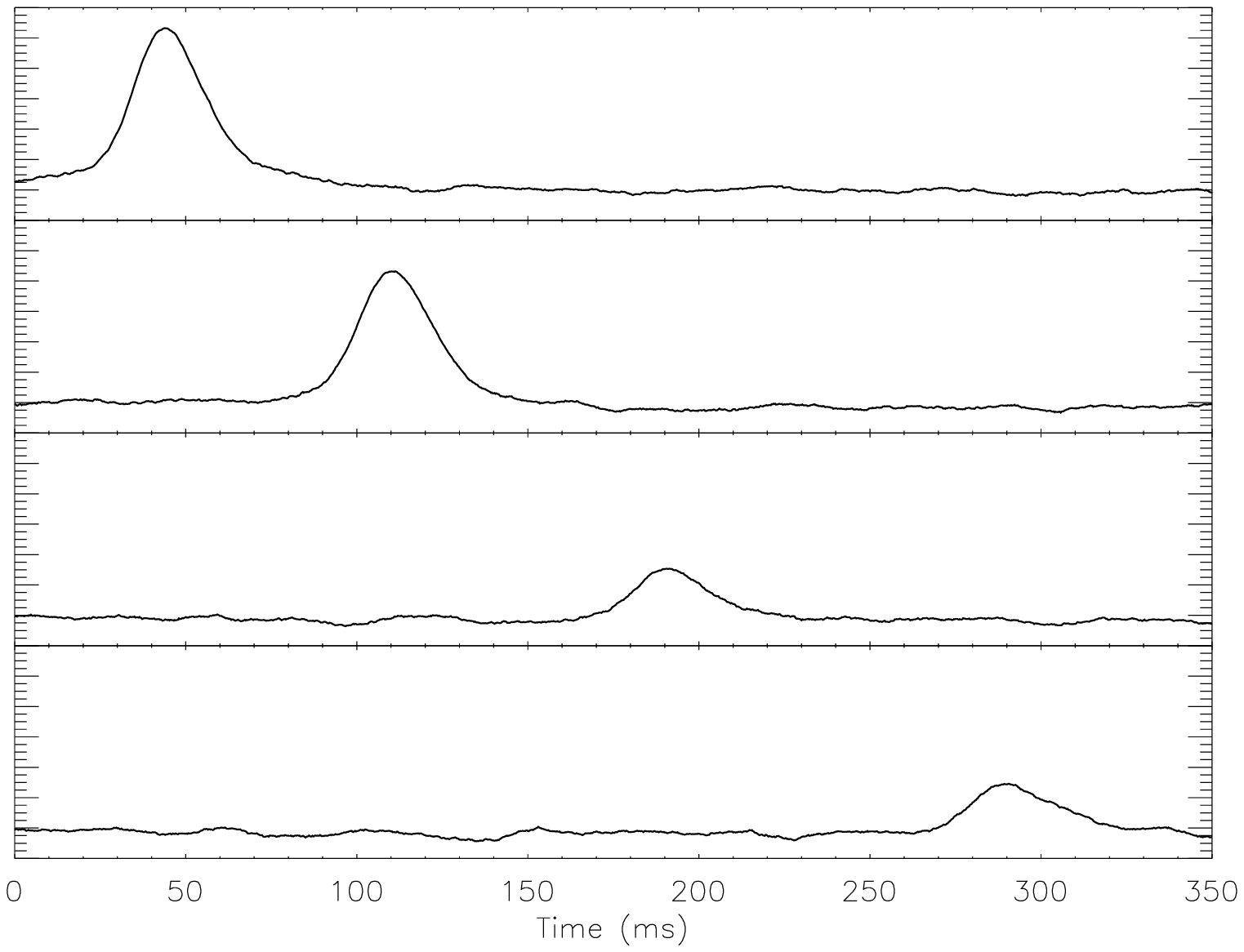}}
\subfigure[Burst on 15th September 2003]{\label{fig:pulseshape_sub3}\includegraphics[width=0.4\textwidth,angle=0]{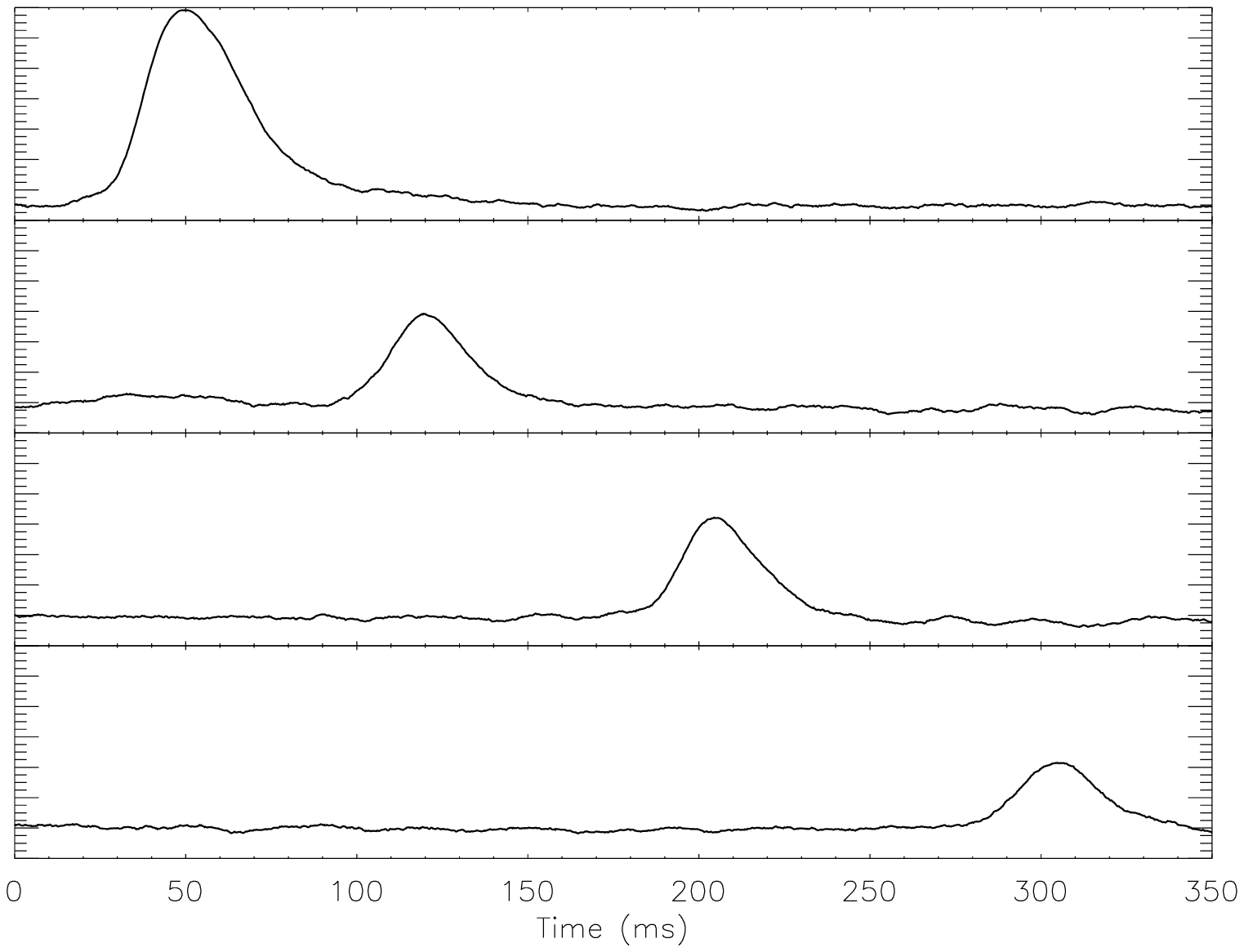}}
  \end{center}
\caption{Pulse shapes in four different subbands for the bursts mentioned in Table \ref{tb:peryton1}. In each case, the signals in all beams have been summed. The central frequencies are 1482, 1410, 1338, and 1266 MHz, from top to bottom.}
\label{fig:all_subpulse}
\end{figure*}

Single-pulse plots\footnote{The pulsar analysis package SIGPROC has been used to make the plots. This package is available online at http://sigproc.sourceforge.net/.}  for the central beam for all of these bursts are shown in Fig. \ref{fig:all_singlepulse}. Dispersive delays for all the bursts are shown in Fig. \ref{fig:allfreqtime}, plotted by summing the signals in all beams and using the mean DM values from all beams. It is clear that although there is no sharp kink at the higher frequency end \citep[as for the sixth peryton of][]{bbemc11}, these are not equally strong over the entire frequency band.This amplitude modulation is not likely caused by interstellar scintillation as the expected scintillation bandwidth ($\sim 1$ kHz) for these bursts, predicted from the latest Galactic electron density model \citep{cl02}, is much smaller than our bandwidth. Therefore, this fact suggests that these are not of celestial origin. Also, the pulses are in general brighter at higher frequencies, resulting in large positive ($6-8$) spectral indices (see Fig. \ref{fig:all_subpulse}). 

For any of these bursts, we did not find any significant variation in the pulse shape in different beams. To check the frequency dependence of the pulse widths, we divided the pulses in each beam into four sub-bands and calculated the pulse widths in each sub-band. We did not find any correlation between pulse widths and the central frequency of the sub-bands, which further supports the conclusion of non-celestial origin. However, we estimated that if these bursts were of celestial origin, they would be located at distances $\sim 7, 6, 7$ and $9$ kpc according to the latest model of Galactic electron density \citep{cl02}.

For all of the bursts, we calculated the peak flux density $S_{\rm peak}$ for each beam. For this purpose, we first calculated the mean ($S_{\rm off,mean}$) and rms ($S_{\rm off,rms}$) fluxes of the off-pulse regions. The radiometer equation predicts rms fluctuations $S_{\rm pred,rms} = \sigma \, \left(S_{\rm sys}+S_{\rm sky}\right) / \sqrt{n_{p} \, t_{\rm samp} \, \Delta \nu} $ where $\sigma = 1.4$ is a factor which accounts for the loss due to one-bit sampling ($\sqrt{\pi/2} = 1.25$) and an additional 15\% loss due to other factors \citep{evb01}, $n_p=2$ is the number of polarization, $\Delta \nu=$ 288 MHz is the bandwidth, $t_{\rm samp}=$ 250 $\mu {\rm s}$ is the sampling time, and $S_{\rm sys}$ is the system temperature\footnote{http://www.atnf.csiro.au/research/multibeam/description.html} (29 Jy for beam 1, 30 Jy for beams 2$-$7, and 36 Jy for beams 8$-$13). Values of $S_{\rm sky}$ have been calculated by scaling the all-sky radio continuum map at 408 MHz \citep{hssw82} with a spectral index of $-$2.6. The calibration constant is $C=S_{\rm pred,rms}/S_{\rm off,rms}$. We calibrated the profiles in Jansky by subtracting $S_{\rm off,mean}$ from the profiles and multiplying by $C$. The values of $S_{\rm peak}$ are reported in Table \ref{tb:peryton1}. These bursts are roughly two times brighter than the brightest peryton detected by \citet{bbemc11}

\begin{table*}
\caption{Parameters for dispersed radio bursts found in the PMPS data. The columns from left to right represent the dates, start times, coordinates of the central beam, values of DM averaged over all beams, peak fluxes averaged over all beams, pulse widths averaged over all beams and brief notes on weather at the Parkes site on the days of the bursts.}
\begin{tabular}{lcccccc} \hline \hline 
 Date & Start Time & RA, DEC &  DM  &  $S_{peak}$  &  $W$  & Weather Summary \\
   &  & of the central beam &  &     &  & \\
   & (AEST)  &(hh:mm:ss, dd:mm:ss)   & (${\rm cm^{-3} \, pc}$)   & (Jy)  & (ms)  & \\ \hline
 3 August 1998 &  12:40:09.264  & 13:23:36, $-$63:02:55   & 421.36  & 0.58  & 23.08 & temperature $6.4 - 15.5^{\circ}$ C,  \\
     & &    &  &     &  &  average wind speed 8 km/h, no rain \\
   &  &  &  &  &     & \\
 12 August 1998 & 12:46:06.696  & 14:42:40, $-$61:15:06   & 356.94  & 0.60  & 24.88 & temperature $6.9 - 16.0^{\circ}$ C,  \\
     & &    &  &  &     &  average wind speed 5 km/h, no rain \\
    &  &  &  &  &     & \\
 12 August 1998 &  12:47:35.976 & 14:42:40, $-$61:15:06   & 393.11  & 0.64  & 28.35 & temperature $6.9 - 16.0^{\circ}$ C,  \\
     & &   &  &  &     &  average wind speed 5 km/h, no rain \\
    &  &  &  &  &     & \\
 15 September 2003 &  13:35:46.872  & 16:10:51, $-$47:06:53    & 374.28  & 0.75  & 30.91 & temperature $2.0 - 17.0^{\circ}$ C,  \\ 
     &  &   &  &  &     &  average wind speed 11 km/h, no rain \\ \hline \hline
\label{tb:peryton1}
\end{tabular}
\end{table*}

\section{Discussions}
\label{sec:disc}

The four bursts we find seem to be of non-celestial origin as they are very bright and visible in all thirteen beams, do not show the expected correlation between pulse width and frequency, and are not equally bright over the entire bandwidth. The question left is whether these were of human-made origin. These four bursts are in general different than other RFI present in the data, which has peak SNR at zero DM most of the time. In other cases, the RFI shows high SNR over a very wide range of DM in the single-pulse plots, but does not show $\delta \, t \propto \nu^{-2}$ relation in the frequency - time plots. Moreover, even when a particular RFI signal shows up in multiple beams, the DM corresponding to the peak SNR is significantly different for different beams. The broad-band nature and emission in the legally protected band (1.4$-$1.427 GHz) for the four bursts suggest against intentional emission. There is a possibility that these bursts are some kind of chirped RFI signal, which in general would appear as a broadband signal. The source of this RFI could be located inside the Parkes site or within the instrument itself. However, \citet{bbemc11} argued that to explain the similar amplitude modulations in all 13 beams in case of the perytons, the source of the RFI must be located at a distance greater than 4~km. The modulation in all 13 beams is similar for our signals, suggesting a distant source following the same logic. But as this argument is applicable only for broad-band emitters, the possibility of on-site origin can not be excluded. 

One of the suggestions by \citet{bbemc11} was the association of perytons with atmospheric events. Therefore we checked the weather at Parkes on the days of the bursts and found that all of these occurred on sunny days unlike perytons (see Table \ref{tb:peryton1}). Also these bursts occurred during August-September, while perytons occurred mostly in late June. However, one similarity is that these bursts occurred close to midday, as did the perytons. This fact supports the possibility of human-made origin of these events as human activity at the Parkes site peaks around midday.

We then investigated any possible associations with  Terrestrial Gamma-Ray Flashes (TGFs). TGFs are very short (lasting up to a few milliseconds) bursts of high-energy photons observed by the BATSE instrument on board the Compton Gamma Ray Observatory (CGRO) \citep{fishman94,nemiroff97}, Reuven Ramaty High Energy Solar Spectroscopy Imager (RHESSI)\citep{grefenstette09}, AGILE satellite \citep{marisaldi10} and Gamma Ray Burst Monitor onboard {\it Fermi}  \citep{briggs10}. TGFs have been associated with strong thunderstorms mostly concentrated in the Earth's equatorial and tropical regions. As thunderstorms are associated with optical flashes (``Sprites'') and very low frequency (kHz) radio bursts (``Sferics''), there is a possibility that lightning produces signals in the GHz band too, at which frequency perytons and these new bursts have been observed. We have searched for any correlations of these new bursts as well as perytons with TGFs. As CGRO and RHESSI were the satellites in operation during the time these events  occurred, we searched CGRO data using NASA's High Energy Astrophysics Science Archive Research Center\footnote{http://heasarc.gsfc.nasa.gov/xamin/SingleBox.html} and RHESSI data using the publicly available list\footnote{http://scipp.ucsc.edu/$\sim$ dsmith/tgflib$\_$public/data/tgflist.txt}. We did not find any concurrency between radio events and TGFs. The closest match is for peryton 14 which occurred on 30th June 2002. A TGF was observed by RHESSI on that day, almost sixteen hours after the occurrence of the peryton and the location of the TGF was near the border of Panama and Colombia. Moreover, the south$-$eastern part of Australia is in general free of TGFs. Thus we conclude against any association between TGFs and perytons and our new bursts. 

Out of the 21 perytons discovered so far, 15 occurred within a five-minute timespan, implying only seven independent events.
Two of our four bursts occurred within only 1.5 minutes, implying three independent events. Given the 14014 hours of data searched by \citet{bbemc11} and \citet{kbb12}, and the 23508 hours of data searched by us, the implied event rates are 9.4~yr$^{-1}$ and 1.1~yr$^{-1}$, respectively. These dramatically different event rates are puzzling, given that the data for both surveys was taken around the same time. In addition, \citet{kbb12} noticed a distinct pattern in the peryton arrival times, as there were several perytons appearing $\sim 22$ seconds after the previous one followed by a shorter and then a longer gap. This temporal behaviour suggested against the association of perytons with weather, as there is no reason for atmospheric phenomena to show such a pattern. Interestingly, we did not see any such pattern in the arrival times of our bursts.

Assuming a homogeneous distribution of radio sources and neglecting the effects of pulse broadening due to scattering in the intergalactic matter, \citet{deneva09} estimated that the original survey of the Magellanic Clouds should have resulted in $10^3$ Lorimer burst-like events above that survey's detection threshold of $0.3$ Jy and the Pulsar Arecibo L-band Feed Array (PALFA) survey should have resulted in around 600 events above that survey's detection threshold of $0.03$ mJy. Consideration of cosmological source population reduces these numbers to 500 and 7 respectively (assuming the same luminosity for all radio bursts). As the PMPS survey was $\sim 4$ times longer duration and used the same multibeam system as the original survey of the Magellanic Clouds, a simple scaling by a factor of 4 of the prediction of \citet{deneva09} leads a prediction of $ \sim 2000$ events with flux $\gtrsim 0.3$ Jy in PMPS data for similar cosmological source population. The effect of pulse broadening due to scattering in the intergalactic matter may lower the expected numbers.
However, it is clear that if the Lorimer burst were a member of a cosmological population we would have detected many of these sources in this survey. This work is the best evidence yet that the Lorimer burst was either a rare, isolated event or due to RFI.

Future improved facilities like SKA might be helpful to detect fainter and broader bursts and we might see extragalactic radio bursts in future long surveys with these improved instruments. Comparison of properties of these newly discovered extragalactic radio bursts with the Lorimer burst might help to understand the origin of the latter. Moreover, understanding the origin of non-celestial bursts like perytons and the new bursts reported in this paper might help us to predict the expected properties of these events in future surveys, which will be useful to distinguish true astrophysical events from this type of events.
 
\section{Conclusions} 
\label{sec:conclu} 

We have discovered four radio bursts in the archival PMPS data. These bursts are neither of celestial origin nor related to any known atmospheric events. However, as we can not relate them to any obvious human-made source, the true nature of these bursts remains unclear. But as these bursts look very similar to perytons, it is unclear whether perytons were indeed related to atmospheric events. It is also noteworthy to mention that we did not find any new Lorimer burst-like event. Also with similar non-detections in other surveys like PALFA, this suggests that the Lorimer burst does not belong to a cosmological source population.

\section*{Acknowledgements}

The authors thank S. Burke-Spolaor, D. R. Lorimer and C. Wilson-Hodge for discussions, M. B. Mickaliger for access to his reprocessed files of PMPS data and the reviewer Matthew Bailes for useful comments on the manuscript. MB and MM are supported by WVEPSCOR. ACN was supported by a NASA Space Grant Fellowship.

\end{document}